%% file: rojo-ichep14-nnpdf30.tex
\documentclass[3p,times,twocolumn]{elsarticle}

\usepackage{ecrc}

\volume{00}

\firstpage{1}

\def\smallfrac#1#2{\hbox{${{#1}\over {#2}}$}}
\newcommand{\be}{\begin{equation}}
\newcommand{\ee}{\end{equation}}
\newcommand{\bea}{\begin{eqnarray}}
\newcommand{\eea}{\end{eqnarray}}
\newcommand{\bi}{\begin{itemize}}
\newcommand{\ei}{\end{itemize}}
\newcommand{\ben}{\begin{enumerate}}
\newcommand{\een}{\end{enumerate}}

\newcommand{\lp}{\left(}
\newcommand{\rp}{\right)}

\usepackage{graphicx}
\usepackage{afterpage}
\usepackage{epsfig}
\usepackage{amssymb}
\usepackage{amsmath}
\usepackage{dsfont}
\usepackage{multirow}

\journalname{Nuclear Physics B Proceedings Supplement}

\runauth{Juan Rojo}

\jid{nuphbp}

\jnltitlelogo{Nuclear Physics B Proceedings Supplement}

\usepackage{amssymb}

\usepackage[figuresright]{rotating}

\begin{document}

\begin{frontmatter}



\dochead{}

\title{Parton Distributions based on a Maximally Consistent Dataset}


\author{Juan Rojo}

\address{Rudolf Peierls Centre for Theoretical Physics, 1 Keble Road,\\ University of Oxford, OX1 3NP Oxford, United Kingdom}

\begin{abstract}
The choice of data that enters a global QCD analysis can have a substantial
impact on the resulting parton distributions and their predictions for
collider observables.
One of the main reasons for this has to do with 
the possible presence of inconsistencies, either internal within an
experiment or external between different experiments.
In order to assess the robustness
of the global fit, different definitions of
a  conservative PDF set,  that is, a PDF
set based on a  maximally consistent dataset, have been introduced.
However, these approaches are typically affected by theory biases
in  the selection of the dataset.
In this contribution, after a brief overview
of recent NNPDF developments, we propose a new, fully objective, definition of a {\it conservative}
PDF set, based on
the Bayesian reweighting approach.
Using the new NNPDF3.0 framework,
we produce various conservative sets, which turn out to be
  mutually in
agreement within the respective PDF uncertainties, as well as with the global
fit.
We explore some of their implications for LHC phenomenology, finding
also good consistency with
the global fit result.
These results  provide a non-trivial validation test of the new NNPDF3.0
fitting methodology, and indicate that possible inconsistencies 
in the fitted dataset do not affect substantially the global fit PDFs.
\end{abstract}

\begin{keyword}


\end{keyword}

\end{frontmatter}


\section{Overview of NNPDF developments}

The accurate determination of the parton distribution functions (PDFs) of
the proton is one of the most important tasks for precision phenomenology at the LHC~\cite{Forte:2013wc}. 
PDFs are one of the limiting factors for the precision of our theoretical predictions for Higgs boson production~\cite{Dittmaier:2012vm}, since their uncertainties degrade the accuracy of the Higgs characterization in terms of its couplings; they induce large uncertainties in high-mass
 New Physics particle production~\cite{Borschensky:2014cia}; and they  affect Standard Model  precision measurements such as the mass of the $W$ boson~\cite{Bozzi:2011ww}.

Until recently, the most updated set from the NNPDF Collaboration was
NNPDF2.3~\cite{Ball:2012cx}, the first PDF to ever include LHC data
from ATLAS, CMS and LHCb. 
The NNPDF2.3 sets have been used in a large number of phenomenological
and experimental studies.
In the benchmarking exercise of~\cite{Ball:2012wy}, NNPDF2.3 was compared
in great detail
to other recent sets including CT10 and MSTW08.

The NNPDF2.3 sets can be accessed though the {\sc\small LHAPDF} library, and they
are also available as internal sets in various widely used codes.
For instance, NNPDF2.3 is one of the internal
PDF sets in the {\sc\small MadGraph5\_aMC@NLO} program
for the automated computation of NLO cross-sections matched
to parton showers~\cite{Alwall:2014hca}.
The leading order version of NNPDF2.3~\cite{Ball:2011uy,Carrazza:2013axa} 
has also been implemented
as internal set in the  {\sc\small Pythia8} Monte 
Carlo event generator~\cite{Sjostrand:2007gs}, where it
has been used as the basis of the
recent Monash 2013 Tune~\cite{Skands:2014pea} of {\sc\small Pythia8}.
As compared to older tunes, the new Monash 2013 tune
 achieves
an improved description of a wide range of collider data, including
the recent forward data from the LHC, thanks partly to the the steeper 
small-$x$ gluon in
NNPDF2.3LO.

In addition to the standard QCD PDF sets, we also recently produced
the NNPDF2.3QED sets~\cite{Ball:2013hta}, which supplement the QCD
DGLAP evolution equations with the corresponding
 QED contributions (see~\cite{Bertone:2013vaa}
and references therein). In addition, we
provided for the first time an unbiased determination of the photon PDF
$\gamma(x,Q^2)$
 from 
experimental data without any theory model assumptions.
A precision determination of the photon PDF is relevant since
photon-induced contributions can be comparable or event dominant
with respect to standard quark-induced production  in a number
of crucial LHC processes like high-mass dilepton production~\cite{Boughezal:2013cwa}
and $WW$ production~\cite{Bierweiler:2013dja}.

On the polarized side, the NNPDFpol1.1 set~\cite{Nocera:2014gqa} was recently released, which
is the first global polarized fit using the NNPDF methodology. 
NNPDFpol1.1 supplements all relevant inclusive polarized DIS data~\cite{Ball:2013lla} with
polarized hadron collider data from the STAR and PHENIX experiments
at RHIC on jet and $W$ boson production.
Remarkably, we are able to find evidence for the first time a non-zero
and positive polarization of the gluon in the proton.\footnote{Consistent
results are found by the DSSV collaboration~\cite{deFlorian:2014yva}.}
The next steps will be using polarized semi-inclusive data to constrain
the quark flavor separation, for which a new
set of fragmentation functions using the NNPDF methodology is
required.

\section{The NNPDF3.0 sets}

NNPDF3.0 is the new release from the NNPDF Collaboration.
It is available in {\sc\small LHAPDF6} since version 6.1.4.
As will be discussed in detail in an forthcoming publication, 
NNPDF3.0  is the first  PDF set fully determined
from a
methodology validated by closure tests.
These closure tests ensure that PDFs determined from pseudo-data
generated  from a known underlying law  reproduce the
statistical distribution of results expected on the basis of the
assumed experimental uncertainties. 
An important consequence is that it can
be demonstrated that
methodological uncertainties are rather smaller that
the standard
theoretical and experimental uncertainties.
The updated NNPDF3.0 fitting
strategy allows to produce PDFs with different
theory inputs and with widely different datasets
with a unique consistent methodology, a robustness that has never
been achieved in the traditional PDF fitting approach.

NNPDF3.0 is based on a global dataset,
that includes all the relevant available experimental
constraints on parton distributions.
The NNPDF2.3 dataset
has been supplemented with
the complete HERA-II deep-inelastic
cross-sections from H1 and ZEUS, the
combined charm production data from HERA, jet production 
from ATLAS and CMS, vector boson rapidity and $p_T$
distributions 
from ATLAS, CMS and LHCb, $W+c$ data from CMS and top quark
total cross sections from ATLAS and CMS.
Some of these new LHC observables provide precious information
on poorly-known PDFs.
For instance, the top quark data provide useful constraints
on the large-$x$ gluon PDF~\cite{Czakon:2013tha,Alekhin:2013nda} and
 $W+c$ data allows to pin down strangeness~\cite{Stirling:2012vh,Chatrchyan:2013mza}.

NNPDF3.0 uses state-of-the-art theoretical calculations
for all collider processes included.
At NLO, all LHC observables are computed without
any approximation using suitable fast interfaces
for NLO calculations: {\sc\small APPLgrid}~\cite{Carli:2010rw}, 
 {\sc\small FastNLO}~\cite{Wobisch:2011ij} and  
{\sc\small aMCfast}~\cite{Bertone:2014zva}  
NLO calculations are then supplemented with NNLO $K$--factors
and electroweak corrections when necessary, using
{\sc\small top++} for top data~\cite{Czakon:2011xx} and 
{\sc\small FEWZ} for electroweak
production~\cite{Gavin:2012sy}.
Jet data is included in the NNLO fits using the improved
 threshold approximation~\cite{deFlorian:2013qia}, validated with the exact
NNLO calculation of the gluon-gluon channel~\cite{Ridder:2013mf}, which allows
to carefully select only those data points with
kinematics for which the threshold approximation is close enough to the
exact calculation~\cite{Carrazza:2014hra}.

In this contribution, I want to focus on a particular aspect of
the NNPDF3.0 analysis, 
namely a new proposal for the definition, in a fully
objective way, a set of parton distributions based on a 
maximally consistent dataset,
in order to explore the possible impact of dataset
inconsistencies in the global fit.

\section{Parton distributions based on a maximally consistent dataset}

The choice of data that enters a global QCD analysis has an important
impact on the resulting parton distributions.
One of the reasons for this is the potential presence of inconsistencies,
either between the various datasets or internal within a given 
experiment.
In order to bypass this problem, and to assess the robustness
of the global fit, various definitions of
a {\it conservative} PDF set have been introduced.
However, these approaches are typically affected by theory bias
since  the selection of a maximally consistent datasets is done
following an expectation of which experiments are
more reliable.
For example, the NNPDF2.3 collider-only fit~\cite{Ball:2012cx} 
is based 
on the expectation that collider data are presumably more
robust than fixed-target data, and the MRST2004 conservative
partons~\cite{Martin:2003sk} excluded various datasets which could
be affected by large perturbative corrections.

Here we propose a new alternative definition of a 
{ conservative} set of 
parton distributions.
The main novelty is removing any theoretical bias that might
affect the data selection of such maximally consistent
dataset, using the tools that the Bayesian reweighting 
framework provides~\cite{Ball:2010gb,Ball:2011gg}.
In this new approach, we start from the NNPDF3.0 NLO and NNLO global fits
with $N_{\rm rep}=1000$ replicas, and compute
for each replica the weight
associated with each individual experiment included in the fit:
\begin{equation}
w_k = 
\frac{(\chi^{2}_k)^{{1\over 2}(n-1)} 
e^{-\frac{1}{2}\chi^{2}_k}}
{\smallfrac{1}{N_{\rm rep}}\sum_{k=1}^{N_{\rm rep}}(\chi^{2}_k)^{{1\over 2}(n-1)}
e^{-\frac{1}{2}\chi^{2}_k}} \, ,
\label{eq:weightsN}
\end{equation}
with $n$ the number of points of this experiment and $\chi^2_k$
is the $t_0$ $\chi^2$ for replica $k$. 

\begin{table}[t]
\begin{center}
\footnotesize
\input{chi2tab_conservative.tex}
\caption{\small Datasets
included in the NNPDF3.0 global fit,  indicating which
of these experiments are included in the conservative
partons fits for the
different values of the
threshold $\alpha_{\rm max}$ that have been used.
 \label{tab:chi2tab_conservative}
}
\end{center}
\end{table}

\begin{figure}[t]
\begin{center}
\epsfig{width=0.365\textwidth,figure=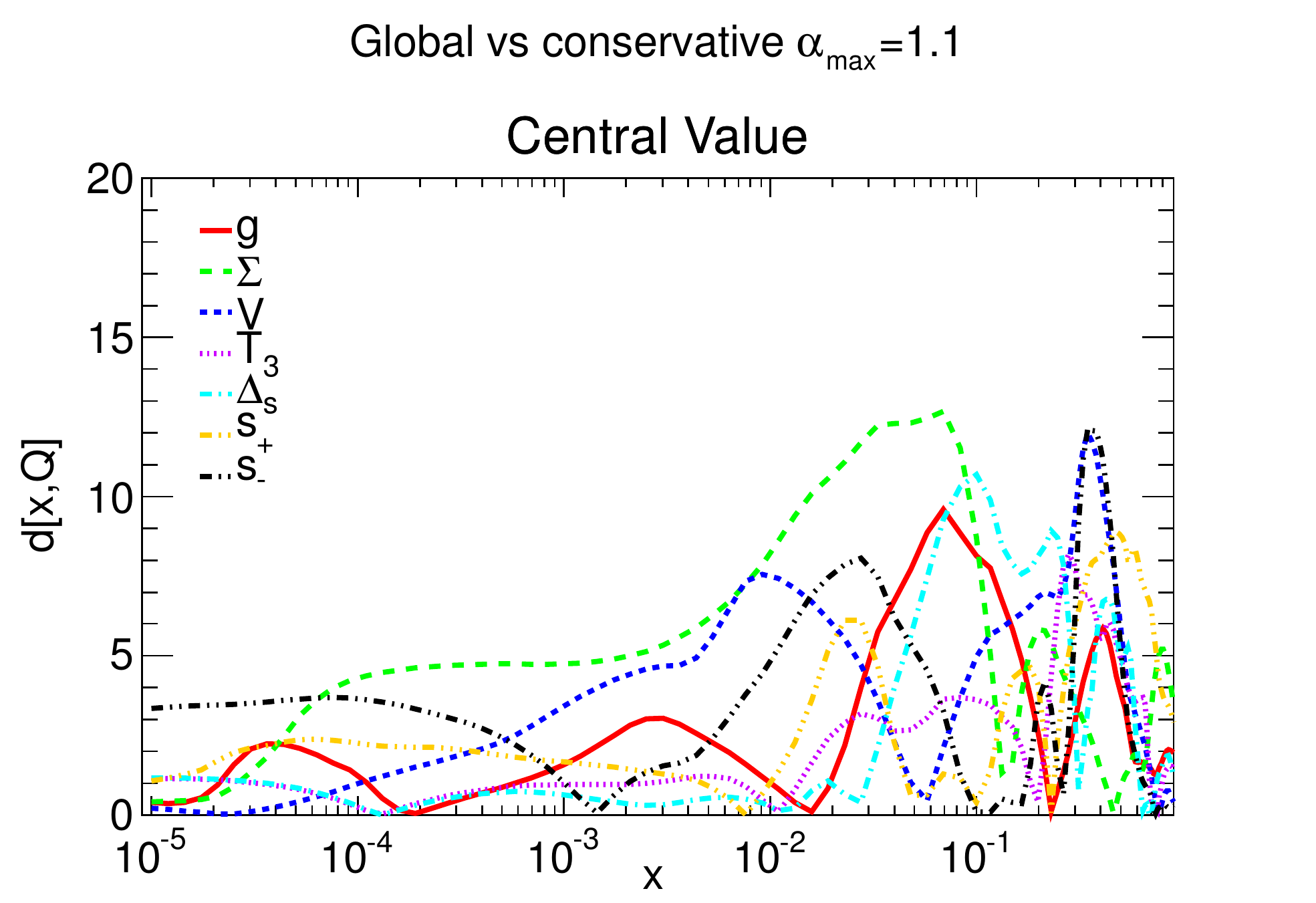}
\epsfig{width=0.365\textwidth,figure=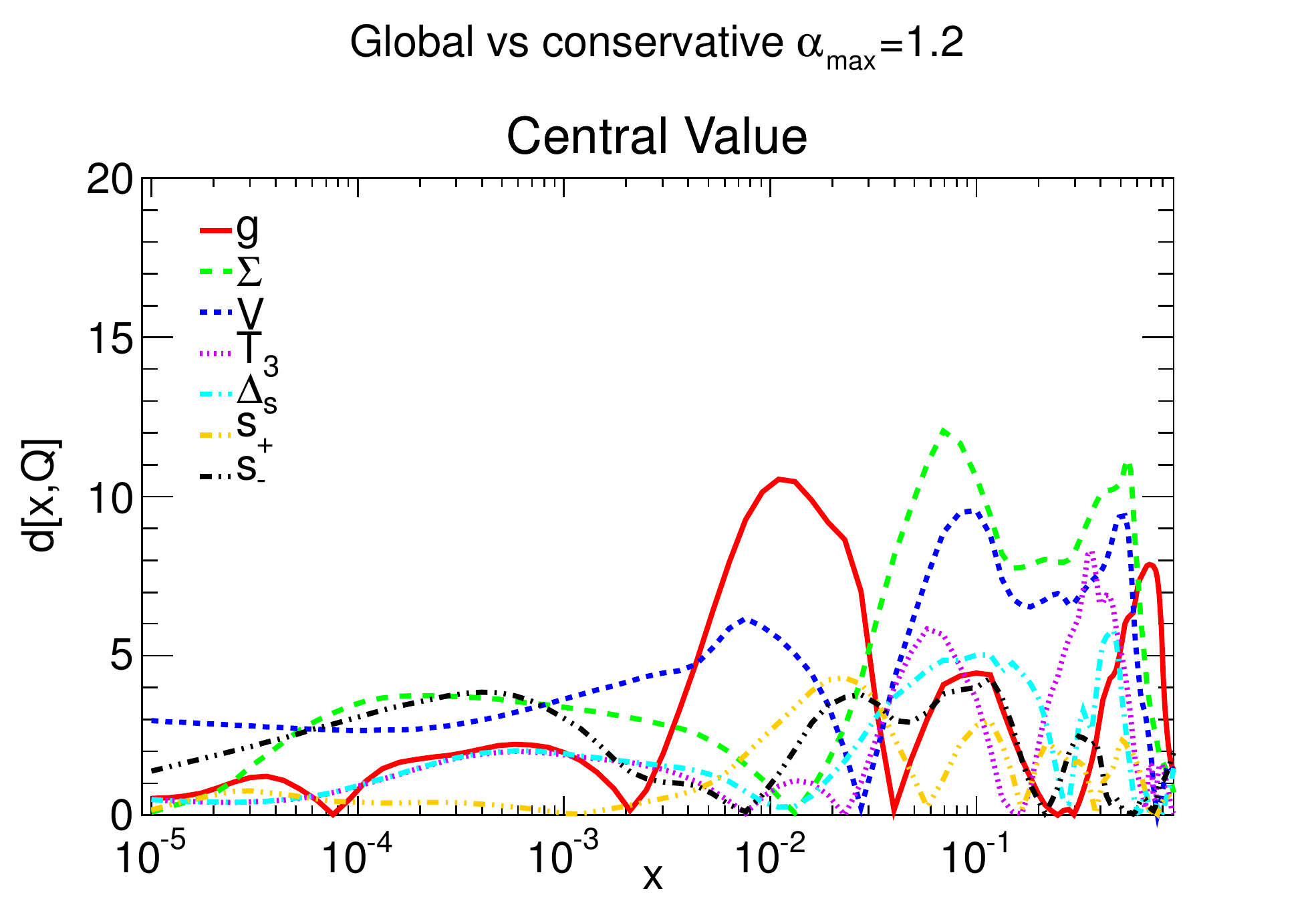}
\epsfig{width=0.365\textwidth,figure=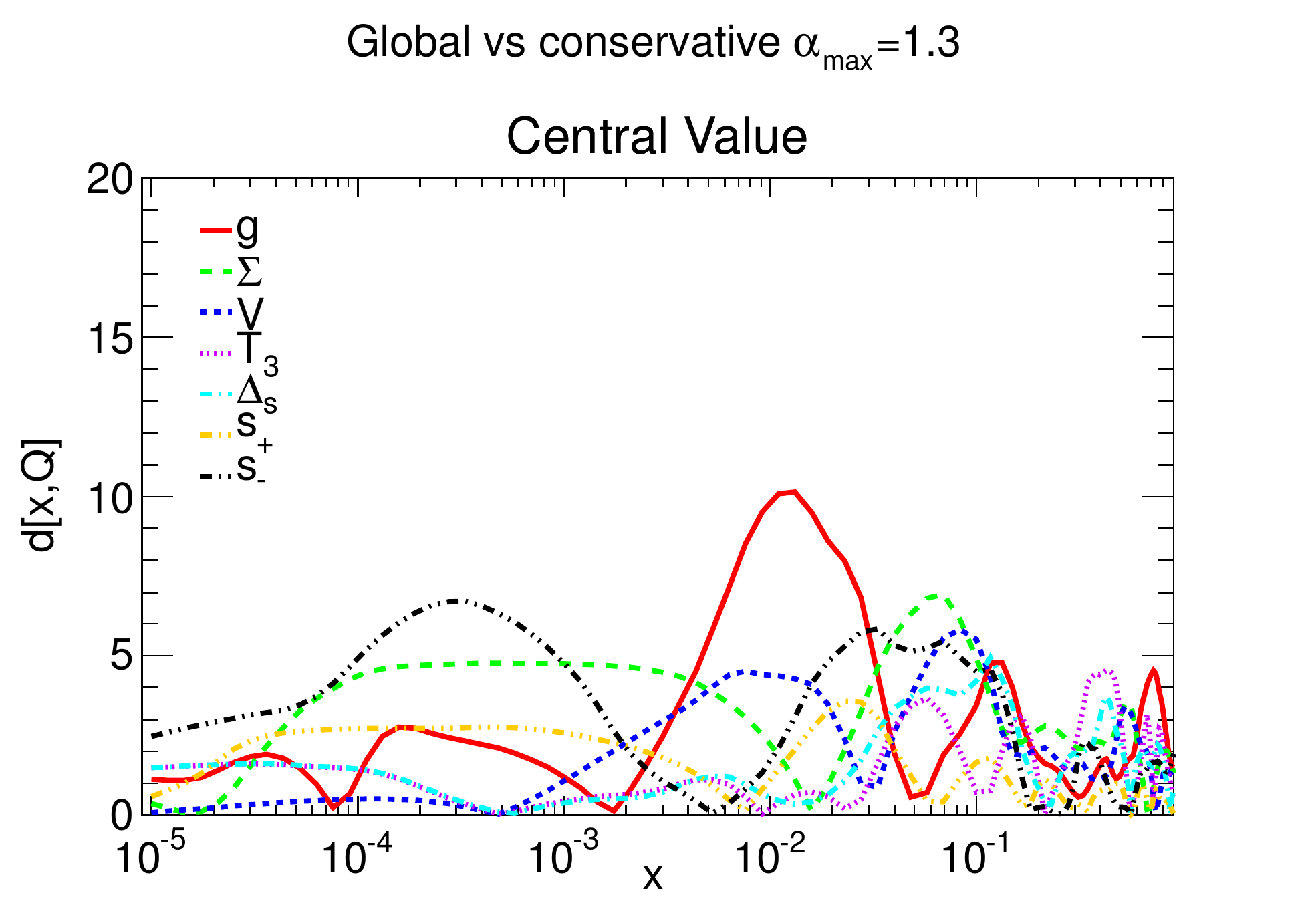}
\caption{\small 
Distances between the global and conservative NNPDF3.0 NNLO partons
for the three values of $\alpha_{\rm max}$ used.
In all these comparisons the sets with $N_{\rm rep}=100$ replicas
have been used.
 \label{fig:distances_global_vs_conservative}} 
\end{center}
\end{figure}

Next, we compute the
$P\lp \alpha\rp$ distributions for each of the individual datasets.
This distribution is a measure of the consistency of the various
experiments with the global fit: the parameter $\alpha$ measures
by how much experimental uncertainties should be rescaled
to achieve perfect consistency.
This distribution is
defined by the fact that when we 
rescale the uncertainties of the data by a factor 
$\alpha$, we can use inverse probability to calculate the probability density 
for the rescaling parameter $\alpha$: 
\begin{equation}
\mathcal{P}(\alpha)\propto \smallfrac{1}{\alpha}\sum_{k=1}^N w_k(\alpha).
\label{eq:rescaling}
\end{equation}
where $w_k(\alpha)$ are the weights Eq.~(\ref{eq:weightsN}) 
evaluated by replacing 
$\chi^2_k$ with $\chi^2_k/\alpha^2$.
When $P\lp \alpha\rp$   peaks close to one, this particular
dataset is consistent with the global fit,
while if it peaks far above one, then it is 
likely that the errors in the data have been underestimated, or
that the theoretical calculations are not accurate enough, for
instance in the presence of large perturbative corrections.

This information suggest an objective criterion to
select a maximally consistent dataset.
For each of the experiment included
in the global fit, we compute the mean, the median and the
mode of the corresponding $P(\alpha)$ distribution.
This experiment will be included in the
conservative set if at least
two out of these three estimators
are below some fixed threshold, denoted by $\alpha_{\rm max}$.
In order to separate effects from different datasets from
perturbative uncertainties, we use the same data at NLO
and NNLO, and we only include in the conservative partons
experiments which satisfy the above criterion both at NLO
and NNLO.
There is some degree of arbitrariness in this criterion,
in particular in the choice of $\alpha_{\rm max}$, 
or by the requirements that NLO and NNLO datasets are the same
in the conservative fits, which is compensated by its
objective character.

The main difference between this and other
criteria is that while previous conservative sets were
 based on an expectation of which data are
more reliable, the new criterion is based on a measure of which data
are consistent or inconsistent with the rest of the experiments
in the global fit.
Again, let me emphasize that this consistency can mean
either consistency of the
new data with the other datasets or
 internal self-consistency.

We have produced various sets of conservative
partons, obtained for three different values
of  $\alpha_{\rm max}$, namely 1.1, 1.2 and 1.3, corresponding
to different degrees of tolerance about the inconsistencies that
we admit in the fitted dataset.
In Table~\ref{tab:chi2tab_conservative}
we list all the datasets
included in the NNPDF3.0 global fit, and indicate which
of these experiments are included in the conservative
partons fit for the values of the
threshold $\alpha_{\rm max}$ that have been used.

In order to gauge the fit quality of the various conservative
PDF sets, in Table~\ref{table-chi2} we show the NLO and NNLO
experimental $\chi^2$ (see~\cite{Ball:2012wy} 
for the precise definition) for
both the global and the three conservative fits.
The results are as expected: by increasing
$\alpha_{\rm max}$ we interpolate between the maximally
consistent dataset with $\chi^2\sim 1$ and the global
fit result.
These results suggest that $\alpha_{\rm max}=1.1$ could be used
to define the best approximation to
a conservative parton set.

\begin{table}[h]
\centering
\begin{tabular}{ccccc}
\hline
& $\chi^2$ NLO & $\chi^2$ NNLO \\
\hline
$\alpha_{\rm max}=1.1$  & 0.97   &  1.01  \\
$\alpha_{\rm max}=1.2$  & 1.06   &  1.09  \\
$\alpha_{\rm max}=1.3$  & 1.12   &  1.15  \\
\hline
Global  &  1.23  &  1.28  \\
\hline
\end{tabular}
\caption{ \small\label{table-chi2} The total $\chi^2$
per data point for
the global and conservative fits for different values
of the threshold $\alpha_{\rm max}$.
}
\end{table}

To quantify how the various conservative partons
different from the global fit, we show in 
Fig.~\ref{fig:distances_global_vs_conservative} 
the corresponding distances for $\alpha_{\rm max}=1.1,1.2$ and $1.3$.
Let us recall that for $N_{\rm rep}=100$, a distance of order 10
correspond to PDFs that agree at the one-sigma level.
As can be seen, for the three fits there is a nice consistency with the
global dataset, with PDFs differing at most at the one-sigma level.
Of course PDF uncertainties are larger in the fits to reduced
datasets, but the statistical compatibility with the global fit
is very similar for three three values of $\alpha_{\rm max}$.

\begin{figure}[t]
\begin{center}
\epsfig{width=0.365\textwidth,figure=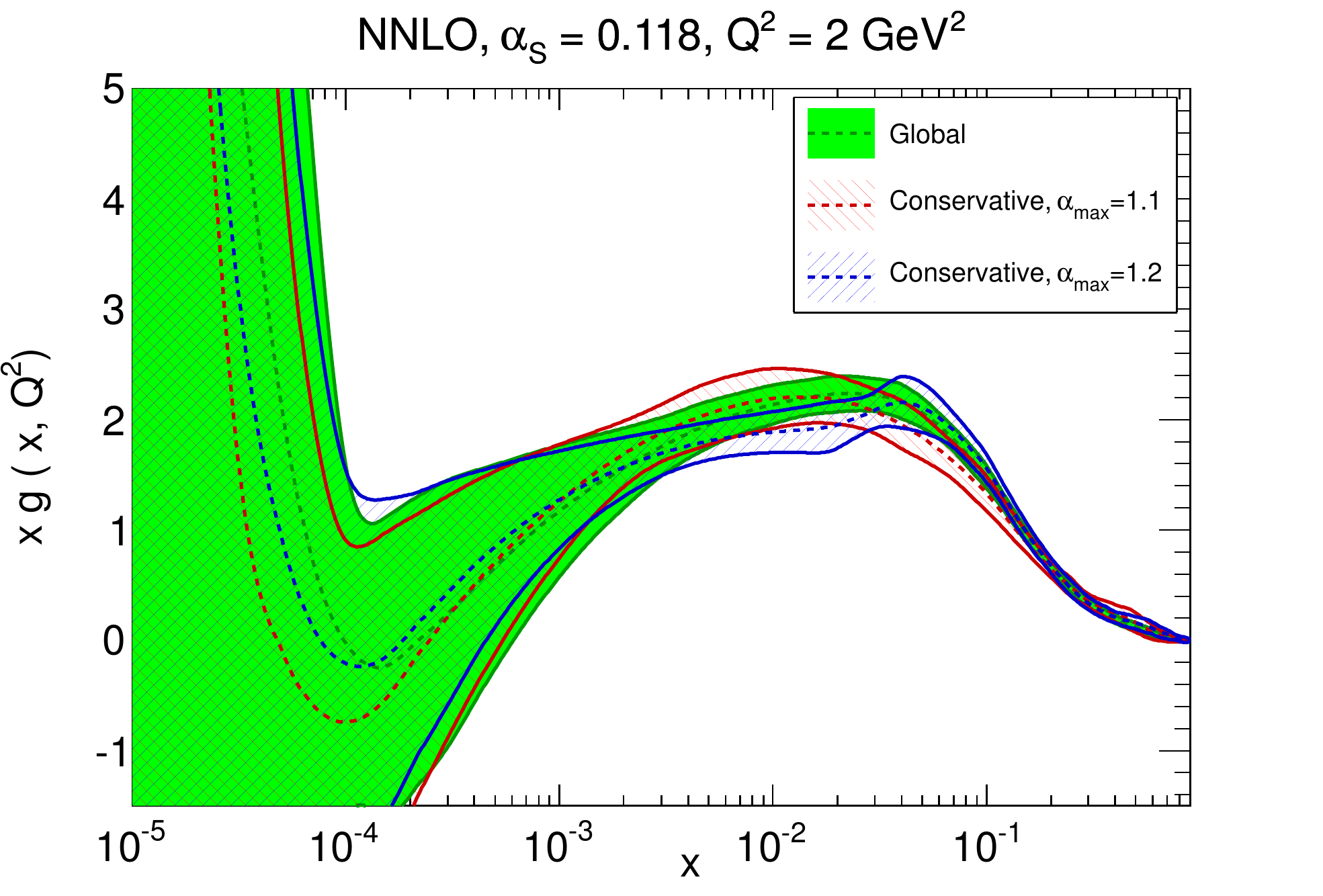}
\epsfig{width=0.365\textwidth,figure=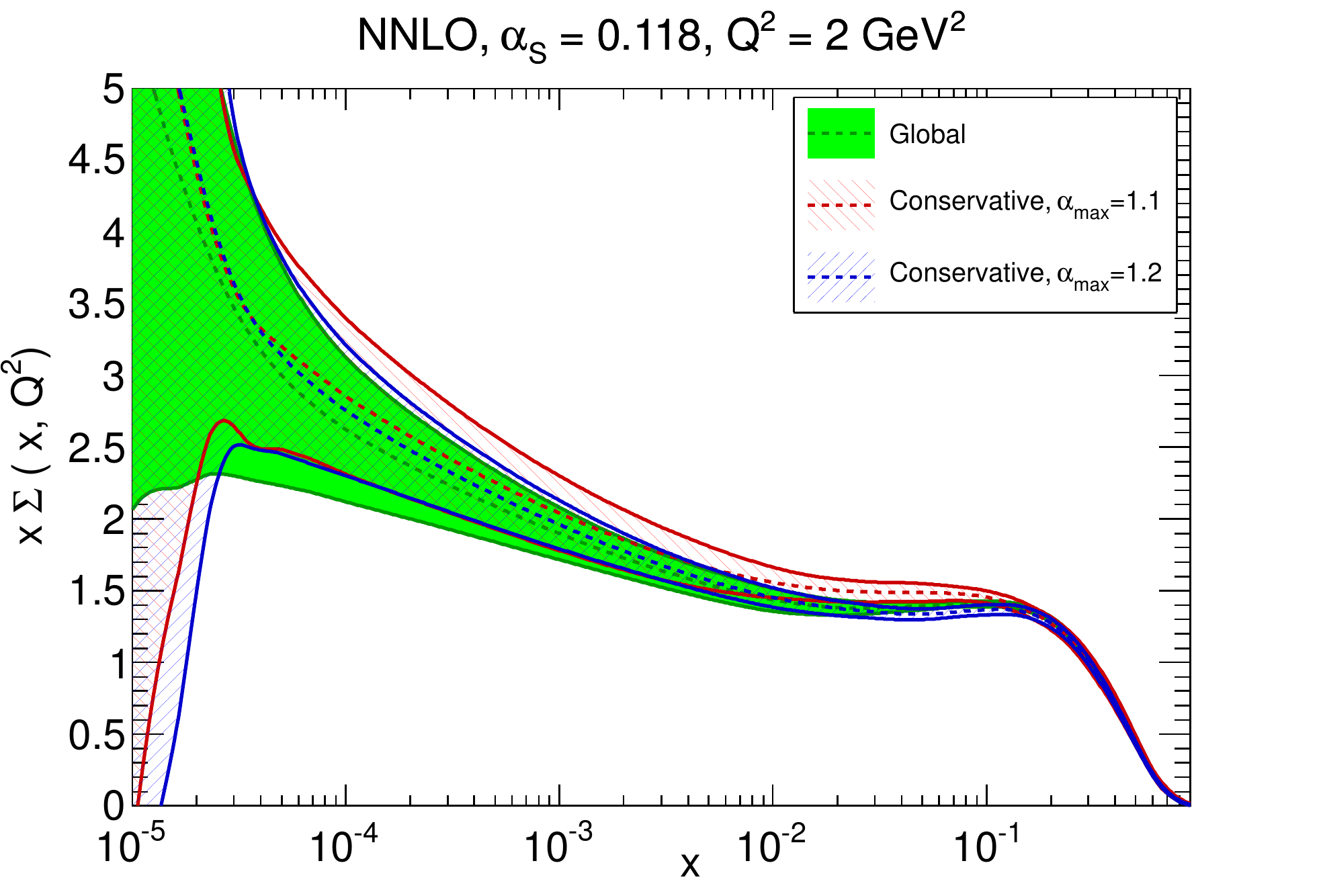}
\epsfig{width=0.365\textwidth,figure=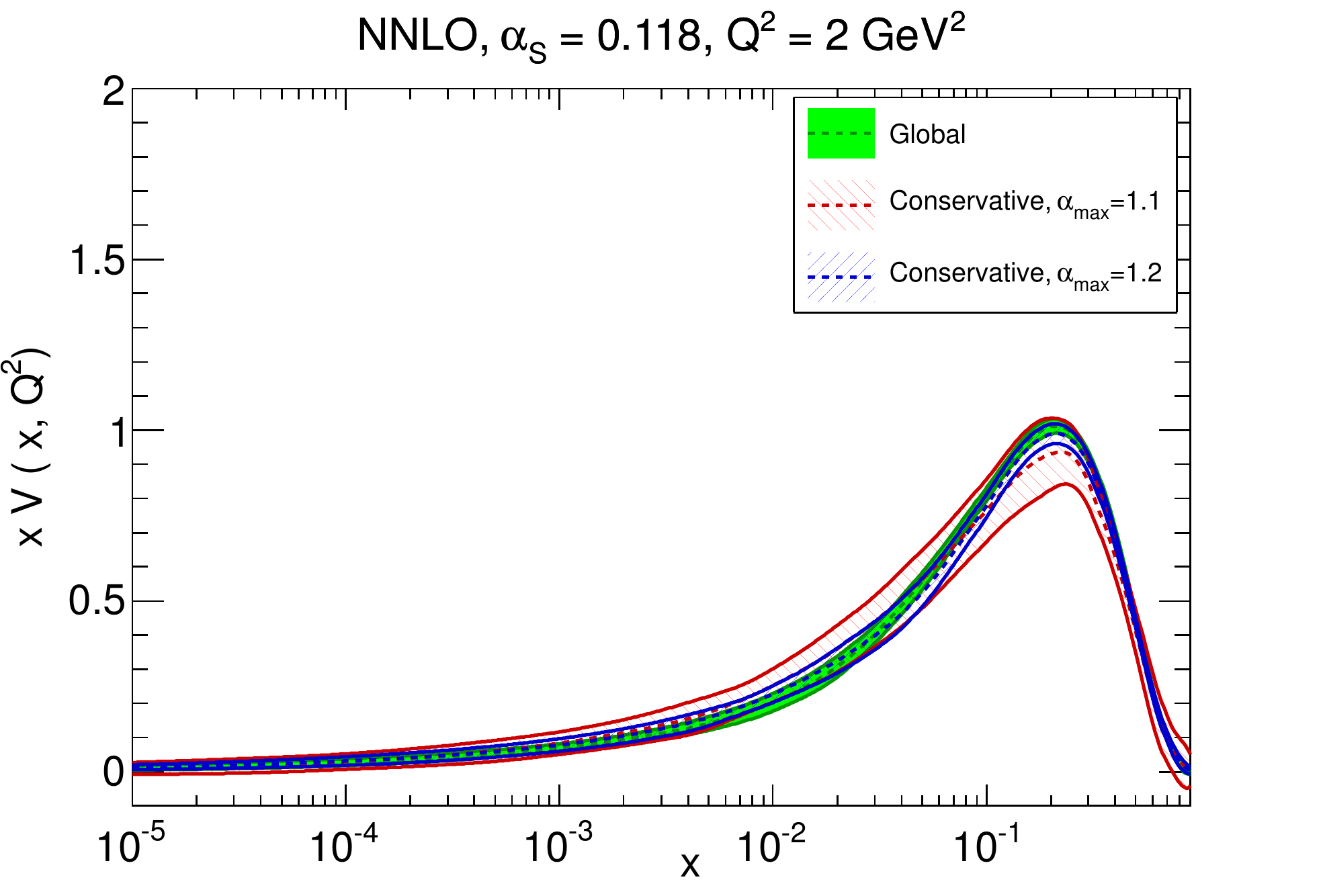}
\caption{\small 
Comparison of various NNLO PDFs in the global fit and
in the two conservative fits with $\alpha_{\rm max}=1.1$ and
1.2.
The comparison is performed at a scale of $Q^2=2$ GeV$^2$.
From top to bottom, we have
gluon, the singlet PDF and the 
total valence PDFs. \label{fig:pdfs_conservative}} 
\end{center}
\end{figure}

Next, we compare various NNLO PDFs in the global fit and
in the two conservative fits with $\alpha_{\rm max}=1.1$ and
1.2, in Fig.~\ref{fig:pdfs_conservative}, at a scale of
$Q^2=2$ GeV$^2$.
As we expected from the distance comparison, 
there is a nice agreement between the different
fits, typically
at the one-sigma level or better.
This is a non-trivial consistency check of both  the global fit approach
and of the definition of { conservative} partons that we propose here.
The small-$x$ gluon is similar in all cases because is driven by the
HERA-I data.
Larger differences are found in the quark sector,
at medium and large-$x$ and for the most conservative
fit with $\alpha_{\rm max}=1.1$.
The region around $x\sim 0.01$ for the gluon, relevant for Higgs production
in gluon fusion, is stable at the one-sigma level, consistent
with the studies based on NNPDF2.3 in the 
Les Houches 2014 proceedings~\cite{Butterworth:2014efa}.

\section{Implications for LHC phenomenology}

\begin{figure}[t]
\begin{center}
\epsfig{width=0.49\textwidth,figure=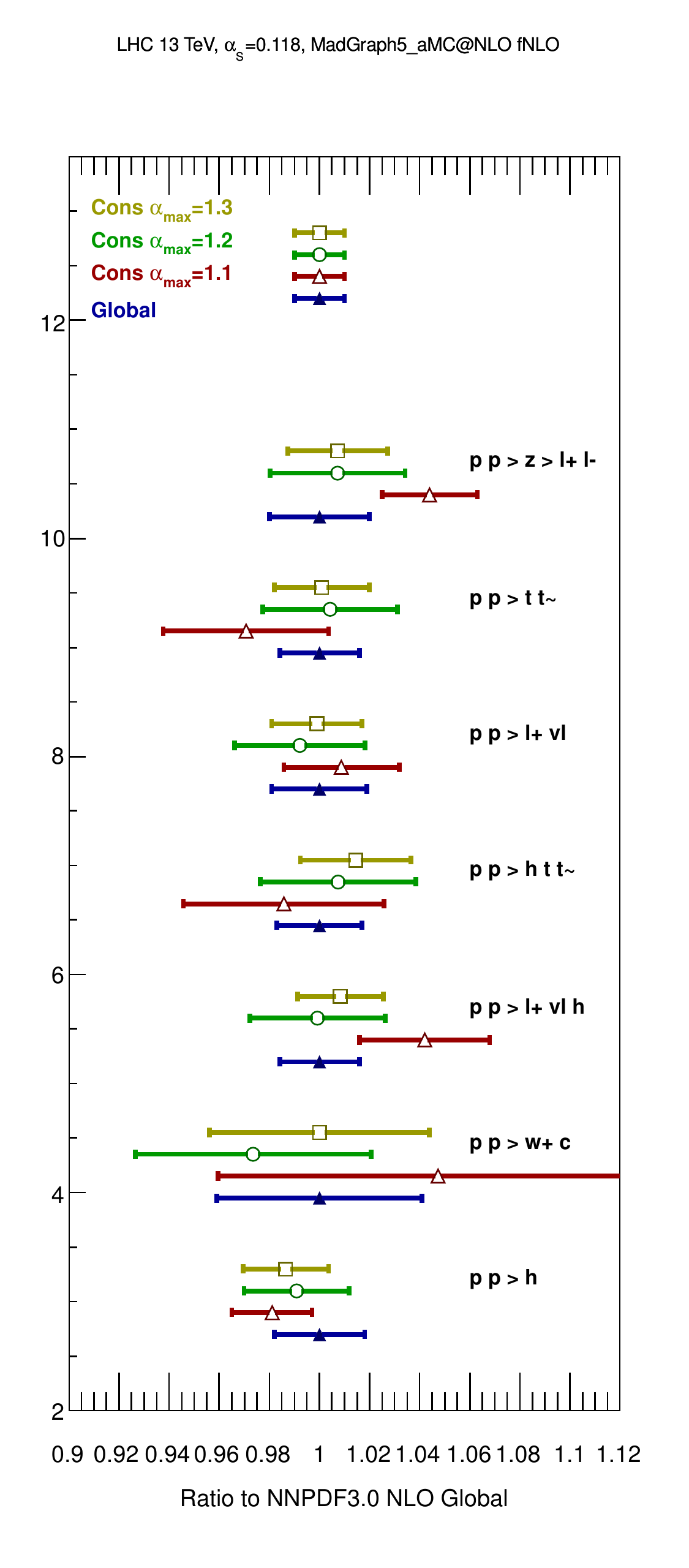}
\vspace{-1.0cm}
\caption{\small 
Comparison of the predictions for a 
number of LHC observables between NNPDF3.0 NLO
and the various corresponding conservative fits.
Results have been computed with {\sc\small MadGraph5\_aMC@NLO} in the
fNLO mode for the LHC 13 TeV and typical LHC analysis cuts.
The cross-sections are shown as ratios with respect to the
NNPDF3.0 NLO central value.
 \label{fig:lhc_conservative}} 
\end{center}
\end{figure}

As we have found, the conservative partons
are nicely consistent with the global fit results for all
values of the threshold $\alpha_{\rm max}$, with of course rather
larger PDF uncertainties, specially for small  $\alpha_{\rm max}$.
To study what are the implications of these conservative
partons for LHC phenomenology, in the following the have used
 {\sc\small MadGraph5\_aMC@NLO}~\cite{Alwall:2014hca} to compute
NLO predictions for a variety of LHC observables both in the global
fit and in the different conservative fit.
Cross-sections are computed at 13 TeV with typical
experimental analysis cuts in the final states.
The results are summarized in Fig.~\ref{fig:lhc_conservative}, 
where we show the predictions for the various processes
as ratios with respect to the NNPDF2.3 NLO global fit predictions.

Is clear that all the fits are consistent at the one-sigma level.
The fits with smaller values of $\alpha_{\rm max}$ have larger
PDF uncertainties and therefore larger fluctuations
of the central values, as expected, while the fit
with $\alpha_{\rm max}=1.3$ is quite close to the global
fit.
The increase in PDF errors is quite noticeable,
for example in cross-sections that depend
on the large-$x$ gluon such as $t\bar{t}$ and
$ht\bar{t}$, as well as for those that depend on strangeness.
On the other hand, the predictions for Higgs production
in gluon fusion are rather stable with respect to the
choice of dataset, as had already been
observed in~\cite{Butterworth:2014efa}.

\section{Outlook}

In this contribution I have presented a new strategy to select
a maximally consistent dataset for a PDF analysis.
We have found that in this new approach there is a good consistency
between fits to reduced datasets and the global fit, at the
level of one-sigma PDF uncertainties or better.
The corresponding predictions for LHC observables are
also in reasonable agreement.
Therefore, these conservative partons can be used in phenomenological
analyses that aim to study how fits based on small but maximally consistent
datasets affect LHC observables.

In addition to the release of conservative partons,
it is conceivable that this method could be used in future
NNPDF releases in order to systematically decide which 
data enters into the global analysis.
The idea would be to systematically remove experiments
from the global fit, compute their $P(\alpha)$ distributions
as discussed above and upon the results decide whether or
not it should be kept in the global dataset.
As more and more data from HERA and the LHC are becoming available,
the question of which data use in the global fits will become
more and more pressing, and the strategy outlined here could
provide a possible way forward.





\input{rojo-ichep14-nnpdf30.bbl}

\end{document}

%% file: chi2tab_conservative.tex
\begin{tabular}{c||c|c|c}
\hline
 & \multicolumn{1}{c|}{$\alpha_{\rm max}=1.1$} &  \multicolumn{1}{c|}{$\alpha_{\rm max}=1.2$} & 
 \multicolumn{1}{c}{$\alpha_{\rm max}=1.3$} \\ 
\hline
\hline
NMC $d/p$ & y & y  & y  \\
NMC  & n & n  & n  \\
SLAC  & n&  n & y \\
BCDMS & n & y  &y  \\
CHORUS  & n& y & y \\
NuTeV  & y & y  &y  \\
HERA-I  & y & y & y \\
ZEUS HERA-II  &n & n & y  \\
H1 HERA-II  & n & n  & n \\
HERA $\sigma_{\rm NC}^{c}$  &n & y &y  \\
\hline
E886 $d/p$  & y& y & y \\
E886 $p$  & n & y & y \\
E605  & y & y & y \\
CDF $Z$ rapidity  &n & n  & n  \\
CDF Run-II $k_t$ jets  & n & y & y  \\
D0  $Z$ rapidity & y & y  & y  \\
\hline
ATLAS $W,Z$ 2010  & n & y  & y  \\
ATLAS 7 TeV jets   &  y &  y & y  \\
ATLAS 2.76 TeV jets  & y & y & y  \\
ATLAS high-mass DY  & n  & n  & n \\
ATLAS $W$ $p_T$  & y & y & y \\
CMS $W$ electron asy  & y & y & y \\
CMS $W$ muon asy   & n &  n & n \\
CMS jets 2011  & y & y & y \\
CMS $W+c$ total  &n & n & n  \\
CMS $W+c$ ratio  & n & n & n  \\
CMS 2D DY 2011  & n &  n &  y \\
LHCb $W,Z$ rapidity  & n & y  & y  \\
$\sigma(t\bar{t})$  & n & n  & n  \\
\hline
\end{tabular}